\documentclass{iaus}
\usepackage{graphicx}

\title[Feedback] 
{Feedback in  Galaxy Formation}

\author[Joseph Silk]   
{Joseph Silk}

\affiliation{Dept. of Physics and Beecroft Institute for Particle Astrophysics and Cosmology,\\ Denys Wilkinson Building, University  of Oxford,  1 Keble Road, Oxford OX1 3RH, UK  \\
and Dept. of Physics and Astronomy, The Johns Hopkins University, 3400 N. Charles St., Baltimore MD 21218, USA \\ 
email: {\tt silk@astro.ox.ac.uk}}

\pubyear{2011}
\volume{277}  
\pagerange{119--126}
\jname{Tracing the Ancestry of Galaxies
                                           (on the Land of our Ancestors)}
\editors{C. Carignan, F. Combes \&  K. Freeman,  eds.}

\def\simlt{\lower.5ex\hbox{$\; \buildrel < \over \sim \;$}}
\def\simgt{\lower.5ex\hbox{$\; \buildrel > \over \sim \;$}}

\begin{document}

\maketitle

\begin{abstract}
I review the outstanding problems in galaxy formation theory, and the role of feedback in resolving them.  I address the efficiency of star formation,  the  galactic star formation rate, and the roles of supernovae and supermassive black holes.
\end{abstract}

\firstsection 
\section{Introduction}
Numerical simulations  of  large-scale structure have met with great success. However these  same simulations fail 
to account for the observed properties of galaxies. 
On large scales, $\sim 0.01-100 \rm Mpc$, the ansatz of cold, weakly interacting dark matter has led to realistic maps of the galaxy distribution, under the assumptions that light traces mass and that the initial conditions are provided by the observed temperature  fluctuations in the cosmic microwave background. On smaller scales, light no longer traces mass because of the complexity of galaxy and star formation. Baryon physics must be added to the simulations in order to produce  realistic galaxies. It is here that the modelling falls apart.

Theory provides the mass function of dark halos. Observation yields the luminosity function of galaxies, usually fit by a Schechter function. Comparison of the two is at first sight disconcerting.  One can calculate the $M/L$ ratio for the two functions to overlap at one point,  for a mass $M^\ast$ corresponding to $L_\ast.$
Define $t_{cool}=\frac{{3\over 2}nkT}{ \Lambda(T)n^2}$ and 
$ t_{dyn}= \frac{3}{ \sqrt 32\pi G \rho}.$  For star formation to occur, cooling is essential, and the condition 
$t_{cool}<t_{dyn}$ guarantees cooling in an inhomogeneous galactic halo where gas clouds collide at the virial velocity.
One finds that 
$$
M_{cool}^\ast=\alpha^{3}\alpha_g^{-2}{m_p\over m_e}{t_{cool}\over t_{dyn}} T^{1+2\beta}.$$
For a cooling function $\Lambda(T)\propto T^\beta,$ over the relevant temperature range ($10^5-10^7$ K), one can take 
$\beta\approx -1/2$ for a low metallicity plasma \cite{Gnat}. The result is that one finds a characteristic galactic halo mass, in terms of fundamental constants, to be of order $10^{12} M_\odot.$ The inferred value of the mass-to-light ratio $M/L$ is similar to that observed for $L_\ast$ galaxies. This is a success for theory: dissipation provides a key ingredient in understanding the stellar masses of galaxies, at least for the ``typical'' galaxy.  
The characteristic galactic mass is understood by the requirement that cooling within a dynamical time is a necessary condition for efficient star formation. 

However
all studies to date produce too many small galaxies,  too many big galaxies in the nearby universe, too few massive galaxies at high redshift, and too many baryons within the galaxy halos. In addition there are structural problems: for example, massive galaxies with thin disks and/or without bulges are missing, and the concentration and cuspiness of cold dark matter is found to be excessive in barred galaxies and in dwarfs.  The resolution to all of these difficulties must lie in feedback. There are various flavours of feedback that span the range of processes including  reionisation at very high redshift, supernova explosions, 
tidal stripping and input from active galactic nuclei. All of these effects no doubt have a role, but we shall see that what is missing is a  robust theory of star formation as well as adequate numerical resolution to properly model the interactions between baryons, dynamics and dark matter.  
 
\section{The luminosity function of galaxies}
\subsection{Low luminosity galaxies}

\begin{figure}[b]
\begin{center}
 \includegraphics[width=3.4in]{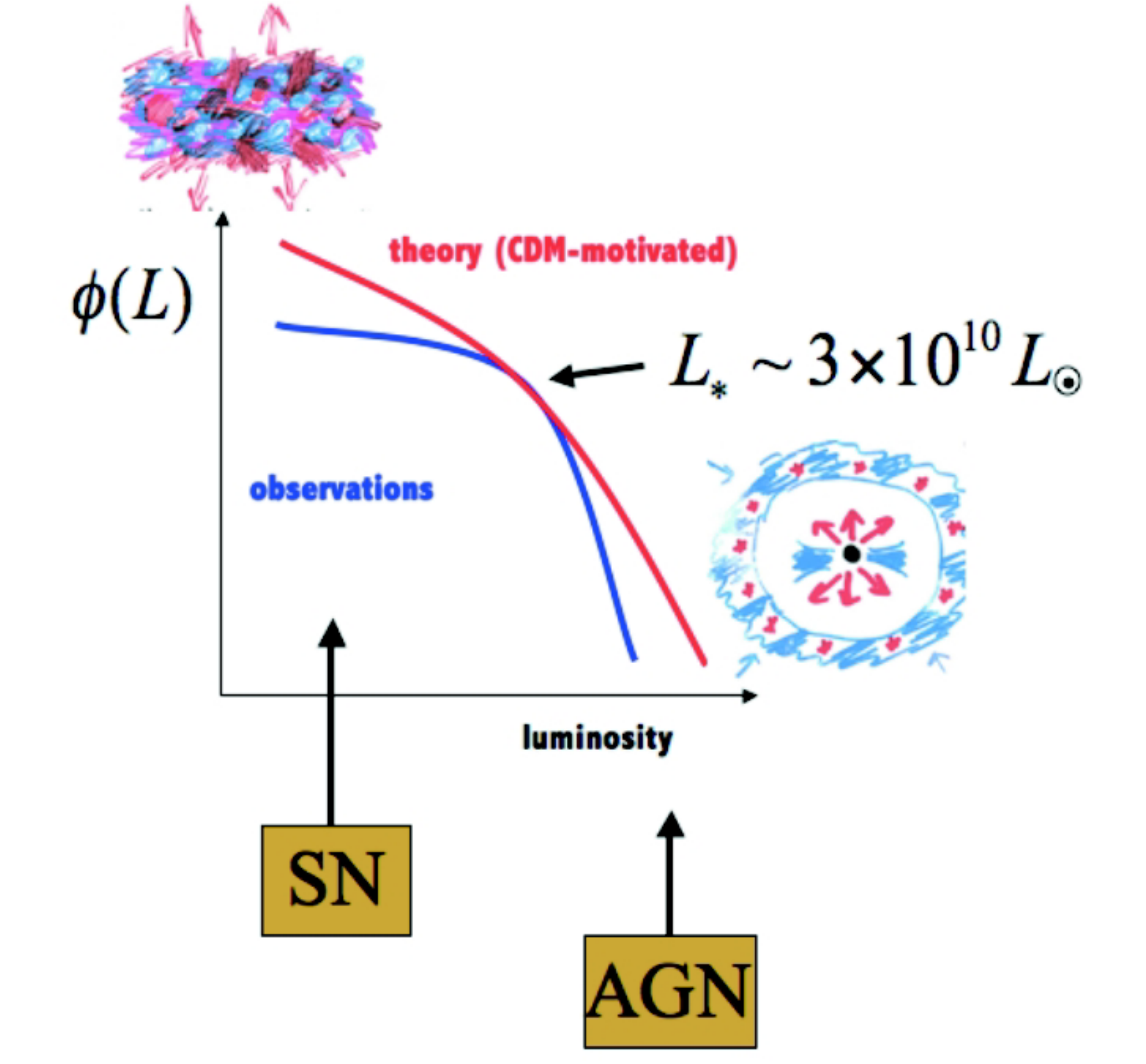} 
 \caption{The theoretical mass function of galaxies compared to the observed luminosity function}
   \label{fig1}
\end{center}
\end{figure}

CDM simulations predict a vast excess of dwarf halos
(Fig.\,\ref{fig1}). 
Input of baryonic physics helps resolve the dwarf excess. Only halos of mass $\simgt 10^5\rm M_\odot$ trap baryons that are able to undergo early $H_2$ cooling and eventually form stars. Reionisation reinforces  this limit by ejecting the baryons in the lowest mass systems that have not collapsed prior to the reionization epoch, and hence suppressing star formation.   Reionisation gives an inevitable feedback for the lowest mass dwarfs. An abrupt increase of the sound speed to $\sim 10-20 \rm km/s$ at $z\sim 10$ means that  dwarfs of mass $\sim 10^6-10^7\rm M_\odot,$ which have not yet collapsed and fragmented into stars, will be disrupted. However more massive dwarfs are unaffected, as are the high $\sigma$ peaks that  develop into early collapsing, but rare, low mass dwarfs. 

The accepted solution for gas disruption and dispersal in  intermediate mass and massive dwarfs  ($\sim 10^8 -10^{10} \rm M_\odot$) is by supernova feedback.  Supernovae expel the remaining baryons in systems of mass up to $\sim 10^8\rm M_\odot, $ leaving behind dim remnants of dwarf galaxies (\cite{dekelsilk}). Presumably the luminous dwarfs accrete gas at later epochs. Most gas is ejected by the first generations of supernovae for systems with escape velocity $\simlt 50 \rm km/s,$ leaving dim stellar remnants behind. Multiphase simulations (\cite{powell11}) confirm the effectiveness of supernova-driven winds in 
 yielding an acceptable fit to the low mass end of the galaxy luminosity function for the classical dwarfs.  A factor of two uncertainty  in abundance remains at intermediate luminosities due to the observed variance at the LMC/SMC level. This is due to  the small number statistics both  locally in the MWG/M31 complex  and observed more widely in the SDSS (\cite{guo11}). The discovery of a population of ultrafaint dwarfs in the MWG halo verifies this expectation, and even provides a quantitative fit to the feedback expectations (\cite{koposov09}) at the ultra-faint end.

%
%
%
%
%

Tidal disruption also plays a role in disrupting satellites whose orbits intersect the disk or bulge.
Dramatic discoveries due to deep imaging of nearby galaxies with very small, wide field of view, telescopes confirm the ubiquity of tidal tails that trace dwarf disruption in the remote past (\cite{martinezdelgado10}). Simulations provide a convincing demonstration that we are seeing tidal disruption in action (\cite{cooper10}).  An independent confirmation of disruption in action comes from studies of the tidal tails around the outermost MWG globular star clusters such as PAL31. Gaps in the tails (\cite {Grillmair2006}) indicate the presence of dark satellites. Numerical simulations (\cite{yoon2010}) find that high M/L satellites of mass $\sim 10^7\rm M_\odot$  are required, again a prediction of the CDM model.

In summary, the low luminosity end of the luminosity function seems to be understood in the context of CDM. However other problems remain unresolved. Most notably, for the dwarf galaxies, there is the question of the apparent persistency of dark matter cores. The simulations prefer cusps, although feedback via, most plausibly, bulk gas motions driven by supernovae can soften the cusps \cite{couchman}.

%
%
%
%

\subsection{Disk galaxies }

In addressing star-forming galaxies, the problem reduces to our fundamental ignorance of star formation. Phenomenology is used to address this gap in our knowledge. Massive star feedback in giant molecular clouds, the seat of most galactic star formation,  implies a star formation efficiency of around 2\%. This is also found to be true globally in the MWG disk, for a star formation efficiency (SFE) defined to be star formation rate/gas mass $\times$ dynamical or disk rotation time.

Remarkably, a similar SFE is found in nearby star-forming disk galaxies. Indeed,  star formation rates per unit area in disk galaxies, both near and far, can be described by a simple law,
with star  formation efficiency being the controlling parameter:
$$SFE =SFR  \times  ROTATION \, TIME  / GAS \, MASS\approx  0.02.$$
The motivation comes from the gravitational instability of cold gas-rich disks, which provides the scaling, although the normalisation depends on feedback physics. For the global law, in terms of star formation rate and gas mass per unit area, supernova regulation provides the observed efficiency of about 2\% which fits essentially all local star--forming galaxies.
One finds from simple momentum conservation  that 
$SFE   =   \frac {\sigma_{ gas}  v_{cool}  m*_{SN}}
{E_{SN}^ {initial}}   \approx   0.02. $ This is a crude estimator of the efficiency of supernova momentum input into the interstellar medium but it reproduces the observed global normalization of the star formation law.

The fit applies not only globally but to star formation complexes in individual galaxies such  as M51 and also  to starburst galaxies. The star formation law is known as the Schmidt-Kennicutt law \cite{kennicutt2006}, and  its application reveals  that  molecular gas is the controlling gas ingredient.  In the outer parts of galaxies, where the molecular fraction is reduced due to the ambient UV radiation field and lower surface density, the star formation rate per unit gas mass also declines
(\cite{bigiel10}).

For disk instabilities to result in cloud formation, followed by cloud agglomeration  and consequent star formation, one also needs to maintain a cold disk by accretion of cold gas. There is ample evidence of a supply of cold gas, for example in the M33 group.
Other spiral galaxies show extensive reservoirs of HI in their outer regions, for example  NGC 6946 \cite{boomsma2008} and UGC 2082  \cite{heald2010}. 
Recent data extends the Schmidt-Kennicutt law to $z\sim 2,$ with
 a tendency for ultraluminous starbursts at $z\sim 2$ to have somewhat higher $SFE$ \cite{genzel2010}.

A more refined  theoretical model needs to take account of star formation in a multi-phase interstellar medium. 
One expects self-regulation to play a role. If the porosity in the form of supernova remnant-driven bubbles is low, there is no venting and the   pressure is enhanced, clouds are  squeezed, and  SN explosions are triggered by massive star formation. This is followed by  high porosity and  blow-out, 
and the turbulent  pressure drops. Eventually   halo infall replenishes the cold gas, the  porosity is lowered and the cycle recommences. Some of this complexity can be seen in numerical simulations \cite{agertz10}. Supernovae provide recirculation and venting of gas into fountains, thereby reducing the SFE and prolonging the duration of star formation in normal disk galaxies.

\subsection{Spheroidal galaxies}
The baryon fraction is far from its primordial value in all systems other than massive galaxy clusters. Supernovae cannot eject significant amounts of gas from massive galaxies. 
Baryons continue to be accreted over a Hubble time and the stellar mass grows. One consequence is  that massive galaxies are overproduced in the models, and that the massive galaxies are also too blue. 

A clue towards a solution for these dilemmas  comes from the accepted explanation of the Magorrian relation, which relates supermassive black hole mass to spheroid velocity dispersion.  This requires  collusion between black hole growth and the initial gas content of the galaxy when the old stellar spheroid formed. One conventionally appeals to outflows from the central black hole  that deliver momentum to the protogalactic gas. When the black hole is sufficiently massive, the Eddington luminosity is high enough that residual  gas is ejected.
An estimate of the available momentum supply come from equating the Eddington momentum with self-gravity on  circumgalactic gas shells, $L_{Edd}/c=GMM_{gas}/r^2.$
Blowout occurs and star formation terminates when the SMBH--$\sigma$    relation saturates. This occurs for $M_{BH}\propto\sigma^{4-5}$, the observed slope  \cite{graham2010}, and gives, at least in order of magnitude, the correct normalisation of the relation. This is the early feedback quasar mode.

There is also a role for AGN feedback at late epochs, when the AGN radio mode drives jets and cocoons that heat halo gas, inhibit cooling, resolve  the galaxy luminosity function bright end problem and account for the red colours of massive early-type galaxies. AGN feedback in the radio mode may also account for the suppression in numbers of intermediate mass and satellite galaxies.  Feedback from AGN in the host galaxies also preheats the halo gas  that otherwise would be captured by satellites.

\section{The AGN-star formation connection}

However reality may be not quite so simple. A more detailed examination suggests that negative feedback  in momentum-driven winds by supermassive black holes falls short of explaining the   observed $M_{BH}-\sigma$  correlation by a factor of a few \cite{silknusser10}.  Moreover comparison of baryonic fractions with bulge-to-disk ratios in nearby galaxies demonstrates that AGN alone do not eject significant amounts of baryons \cite{Anderson}. 
If negative feedback  in momentum-driven winds by supermassive black holes cannot explain  the $M_{BH}-\sigma$   correlation, something else is needed  

\subsection{Its not SN, its not AGN: maybe its both! }
A plausible addition to the physics is inclusion of star formation, induced, enhanced and quenched by the SMBH outflows. There is extensive evidence, recently compiled by Netzer \cite{netzer10}, that demonstrates the intimate connection of AGN luminosity and star formation rate over a wide dynamic range. 
 If  AGN-driven outflows  trigger star formation, the star formation rate is boosted by a factor $t_{dyn}/t_{jet},$ and the outflow momentum is amplified by supernovae \cite{norman}. The star formation rate boost factor amounts to $ ~ v_{cocoon}/ \sigma   \sim 10-100.$ The outflow momentum is amplified by supernovae. Consequently, the momentum supplied to the gas is boosted by the combination of  AGN and star formation.
Of course the causal direction is uncertain, and indeed the phenomena could be mutually self-regulating. To go beyond phenomenology, many details need to be refined, the most pressing perhaps being the nature of the  black hole growth. 

\subsection{Triggering is observed}  
There are examples of jet-induced global star formation, as seen locally  in Minkowski's object \cite{croft06}, and jet-induced CO formation (and excitation) at high redshift. CO is a prerequisite for star formation, and has been detected in large amounts in the host galaxies of high redshift quasars. The conversion ratio to $H_2$ is uncertain however, and renders any conclusions uncertain. The SFE in early type galaxies, possible sites of AGN,  containing CO is elevated \cite{wei2010}. An actively accreting massive black hole in the dwarf starburst galaxy Henize 2-10 \cite {reines2010} is suggestive of current epoch triggering.
At high redshift \cite{wang2010}, the presence of a SMBH favours the 
 higher SFE  seen in ULIRGs.

\subsection{The role of AGN at high $z$}
Are active galactic nuclei  aftermaths or precursors to star formation?  Most data points to a relative increase of black hole mass with redshift. Contrary to earlier claims, even  SMGs, typically the most massive galaxies at $z\sim 2$ in which  the star formation rate is high, reveal black hole mass to bulge mass ratios that agree with the local value\cite{hainline2010}. The most extreme case is  that SMBH  in $z \sim 6$ quasars lie high  \cite{
Riechers2010}. 
The initiation of SMBH growth remains a mystery. The observed  accretion   rate    is   $\sim 10^{-3}$  of the star formation rate in most AGN. The accretion rate tracks the star formation rate at $z\simlt 2$ but any conclusion about the ratio at higher $z$ is confused by the fact that up to 80\% of ultraluminous infrared starbursts (ULIRGs) have buried AGN \cite{imanishi2010}.
AGN are generally thought to be responsible for quenching of star formation, especially at high $z$, rendering massive ellipticals red. However the preceding discussion suggests that there may have been a prior, short-lived, phase of  triggering. The various correlations between AGN content via emission line diagnostics  and  stellar population content and age are inconclusive in elucidating this issue \cite{schawinski}.

                   \section{ Star formation and gas accretion}

Star formation seems to be too complex to be simply gravity-induced. Merging and AGN triggering  are culprits for playing possible roles. What seems to be progressively clear is that there are two distinct modes of star formation. One mode occurs without any intervention from active  galactic nuclei and is characteristic of disk galaxies such as the MWG, on a time-scale of order  at least several  galactic rotation times. Another mode is more intense, occurring on a relatively rapid time-scale, and involves the intervention of AGN, at least for quenching and possibly for  enhancement or even triggering. 

The most important aspect of star formation is the role of the raw material, cold gas. There are  
two modes of gas accretion, which may be classified as cold flows/minor mergers  and     
 major mergers/cooling flows The cold flows occur  in filamentary streams that follow the cosmic web of large-scale structure, and include minor mergers via the dwarf galaxies that similarly trace the web \cite{dekel2009}. Theory suggests that at low redshift, gas accretion by cold streams is important, and that the cold streams are invariably clumpy and essentially indistinguishable from minor mergers of gas-rich dwarfs. Major galaxy mergers account for the observed morphological distortions that are more common at high $z$ \cite{toomre} and generally lead to cloud agglomeration,  angular momentum loss  and cooling flows that feed star formation  \cite{bourneaud}. 

Observationally, one finds  that cold flows are rarely observed. This is because of the small covering factor of the filaments (\cite{Stewart2010, 
Faucher-GiguerKeres2010}). Indirect evidence in favour of cold accretion comes from studies of star formation in dwarfs. The best example may be the Carina dwarf where three distinct episodes of star formation are found \cite{TolstoyHill2009}. However at high redshift, major mergers between galaxies are common. Indeed  ULIRGs are invariably undergoing major gas-rich mergers and dominate the  cosmic star formation rate history at $z\simgt 2,$  whereas normal star-forming galaxies predominate at low redshift ($z \simlt  2$) \cite{leborgne2009}.  This certainly favours the idea of massive spheroid formation by major mergers. 

A  recent compilation \cite{Gonzalez09} of the specific star formation rate (SSR, or star formation rate per unit stellar mass) to $z\sim 7$ in the GOODS field suggests that the star formation time-scale (or 1/SSR) goes from the MWG value of $\sim 10$Gyr at low redshift to $\sim 0.5$Gyr at $z\simgt 2.$ This results provides the primary argument for two distinct feedback-regulated modes of star formation: at low redshift via supernovae and without  AGN, and at high redshift  with, most plausibly,  quenching and possibly  triggering by AGN playing a central role. One would expect a transition between these two modes as the AGN duty cycle becomes shorter beyond $z\sim 1.$ 
A related triggering mechanism appeals to enhanced merging  at high $z$ \cite{khochfar2011}. Alternatively, it has been argued that intensified  halo  cold gas accretion at early epochs may account for all but the most  the extreme star formation rates  at high $z$, although this may require an  implausibly high SFE
\cite{dekel2009}.

\begin{figure}[b]
\begin{center}
 \includegraphics[width=3.4in]{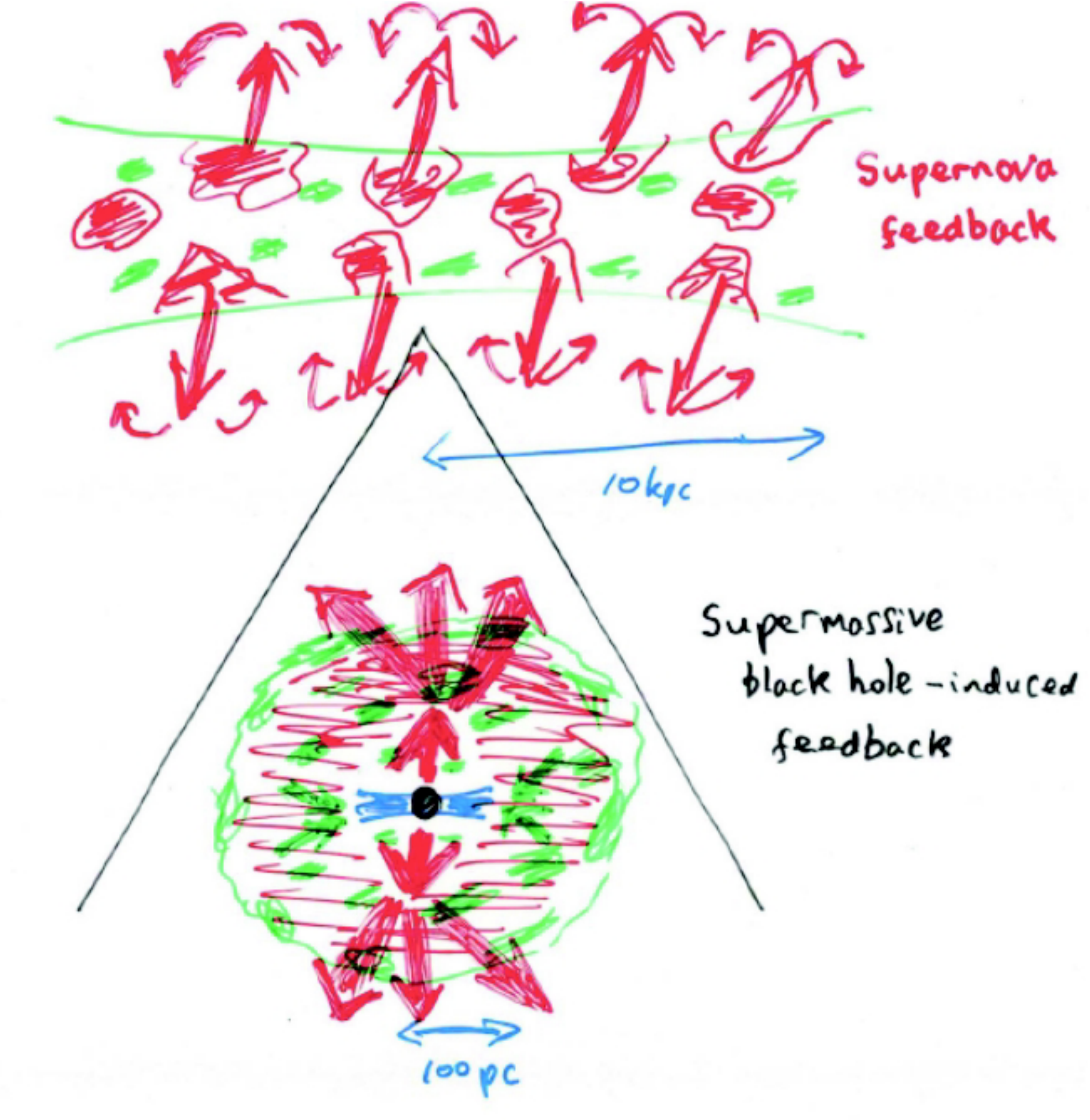} 
 \caption{The rationale for two modes of star formation}
   \label{fig2}
\end{center}
\end{figure}

If the  disk formation mode is  indeed distinct  from the spheroid formation mode, then SMBH might be expected to show some reflection of  alternative growth histories. So-called pseudobulges form from secular instability of disks and contain smaller SMBH than do the more  massive bulges that may have formed via major gas-rich mergers. It is interesting that
SMBH  in pseudobulges lie low on the Magorrian relation \cite{pseudobulges}, as do SMBH in disks relative to those in ellipticals
\cite{graham2010}. These results are for the local universe. 
Recent data on $z\sim 6$ quasars suggest that the most massive black holes also lie high on the black hole/dynamical mass relation \cite{wang2010}.
 Much work still needs to be done to see whether allowance for two modes of star formation can help resolve some of the outstanding problems in galaxy formation (Fig.\,\ref{fig2}).
In addition to the many uncertainties in star formation theory (and I have not addressed one of the key issues, that of the IMF), there remains the nature of black hole growth.  Whether the black holes grow by gas accretion, in which case feedback may play a role in angular momentum transfer \cite{vincenzo}, or by mergers, or by an appropriate  combination, remains unresolved.

\subsection{Galaxies downsize} 
Our understanding of galaxy formation is driven by observations. A good example of this is the phenomenon of downsizing.
This was not anticipated by theorists. Prior to 2000 or so, it was accepted that hierarchical galaxy formation predicted that small galaxies form prior to massive galaxies. Moreover the dynamical or collapse  time of a newly condensed protogalaxy increases with epoch and hence mass. Hence star formation time should likewise increase with mass, if star formation time tracks free-fall time. 

The first indications that this was in error came from the recognition that $[\alpha]/[Fe]$ metallicity ratios were systematically enhanced for the more massive early-type galaxies, indicative of a shorter star formation time. In effect, we have a  cosmic clock: incorporation  into stars of debris
 from SNII ($ \simlt 10^8$  yr)  versus  SNI ($\simgt 10^{9}$ yr) provides a means of dating the duration of star formation.
This result was soon followed by infrared observations that showed that stellar mass assembly favoured more massive systems at earlier epochs. Even metallicity, via  the $ [O/Fe]$  ratio, has been found to demonstrate downsizing. 

Clearly, baryon physics is far more complicated than assumed in the early models of the 1990s. In fact we stlll lack an adequate explanation. Attempts to patch up the problem at low redshift, to avoid an excess of massive galaxies,  exacerbate the inadequacy of the predicted numbers of massive galaxies at high redshift \cite{fontanot2009}. One attempt to correct the problem at large redshift incorporates for the first time thermally pulsing  AGB  (or carbon) stars in  the models, and the extra NIR luminosity reduces the inferred galaxy masses \cite{henriques2010}.
However the price is that the lower  redshift galaxy count predictions no longer fit the models. 
A clue as to the nature of a possible solution may come from the fact that quasars also reveal luminosity downsizing.  This translates into downsizing of central supermassive black hole mass. One might be able to connect the two phenomena if  feedback from AGN were initially positive and also a strongly nonlinear function of SMBH mass.


\subsection{Semi-analytical models of galaxy formation}

Given current computational constraints,  it is impossible to achieve the sub-parsec or finer  resolution needed to adequately model star formation in a cosmological simulation. Theorists have invented a swindle, wherein the complex processes of srar formation are  hidden inside a black box called 'sub-grid physics" that can be tagged onto to a large-scale simulation. This results in semi-analytical models of galaxy formation (SAMs)  which have been remarkably successful in constructing mock catalogues of galaxies at different epochs and are used in motivating and in interpreting the large surveys of galaxies.

However attempts to solve the problems of high redshift galaxies have so far been woefully inadequate.
For example, the early SAM feedback models used AGN quenching, and required excessive dust in early types in the nearby universe \cite{bower2006}. Refinements to high redshift attempted to account simultaneously for galaxy and AGN accounts, and only succeeded by requiring inordinate amounts of dust in order to hide most of the AGNs seen in  deep x-ray surveys
\cite{fanidakis2010}.
A frank assessment of the state-of-the-art in SAMs \cite{benson2010} found that there were some 70 free parameters that need to be adjusted in the newest SAM. An early indication that SAMs were entering uncertain territory can be seen in the early predictions of the cosmic star formation history \cite{springelhernquist2003}.
This showed that as numerical resolution was increased, the predicted star formation rate increased without limit. This makes one  begin to doubt the predictive power of SAMs. 


\section{SUMMARY}

There are a number of inadequately explained phenomena in addition to downsizing.
We do not understand the high frequency of  giant disk  galaxies that are  bulgeless
\cite{kormendy2010}. Many of these are massive disks. The problem here is that early merging, intrinsic to hierarchical formation, results in torquing and angular momentum loss. Not only are bulges inevitable, but essentially all models,  with any combination of cold stream/minor merger/major merger scenarios for gas delivery to drive both star formation and SMBH feeding, produce overly massive bulges compared to the disk masses. The fundamental plane of spheroids displays remarkably little scatter \cite{zaritsky2010}. This is a challenge for SAMs in which merging plays an important role. As for SMBH, we lack any theory of their formation, or perhaps more accurately, we have too many theories.
Star formation remains the outstanding problem. We have no fundamental theory of how stars form.  The early dream of Eddington,
 {\it  ``imagine a physicist calculating on a cloud-bound planet and ending with the dramatic conclusion, 
 `What happens is the stars' "} has not so far been fulfilled in any quantitative sense, such as deriving the IMF,  even though it remains qualitatively correct. 
We are tempted to extrapolate star formation laws, as measured locally, to the extreme conditions found in major mergers. There are already indications that this extrapolation may be premature. But we have no robust alternative.
The one ray of hope is that physics was simpler when the first stars formed, when there were no heavy elements, no dust, and no magnetic fields. This makes computations almost tractable.
The crucial ingredient in star formation is the baryons. But we cannot fully account for the discrepancy between the initial baryon abundance and that observed in galaxies. Ejection  certainly occurs, as evidenced by the enrichment of the IGM, but the details  remain elusive.

Perhaps the greatest uncertainty lies in the nature of star formation. Is the star formation mechanism the same in all environments and at all epochs? Is the IMF universal?
Improved resolution in theory and observation is needed. The great projects of the future, including the ELTs, JWST and LSST, will surely play key roles in this endeavour.
%

{}

\end{document}